\documentclass[runningheads]{llncs}
\usepackage{graphicx}
\usepackage{subcaption}
\usepackage{hyperref}
\usepackage{booktabs}
\usepackage[dvipsnames]{xcolor}

\newcommand{\cands}{\mathcal{C}}

\newcommand{\election}{\mathcal{E}}
\newcommand{\ballots}{\mathcal{B}}

\newcommand{\seats}{N}

\newcommand{\quota}{Q}
\newcommand{\pluseq}{\mathrel{+}=}

    {\begin{center}\begin{minipage}{\textwidth}\begin{tabbing}}%
    {\end{tabbing}\end{minipage}\end{center}}

\newcommand{\ignore}[1]{}

\begin{document}
\title{Analysing Guarantees in Australian Senate Outcomes}
%
%\titlerunning{Abbreviated paper title}
% If the paper title is too long for the running head, you can set
% an abbreviated paper title here
%
\author{Michelle Blom}
\authorrunning{}
% First names are abbreviated in the running head.
% If there are more than two authors, 'et al.' is used.
%
\institute{School of Computing and Information Systems \\
The University of Melbourne\\
Parkville, Australia \\
\email{michelle.blom@unimelb.edu.au}}
\maketitle              % typeset the header of the contribution
\begin{abstract}
Single Transferable Vote (STV) is used to elect candidates to the 76 seat Australian Senate across six states and two territories. These eight STV contests are counted using a combination of ballot scanners, manual data entry and tabulation software.
On election night, some properties of the set of cast ballots are determined by hand. This includes the first-preference tallies of each party. This technical report considers whether there are some properties, such as individual candidates' first preference tallies, that, if assumed to be accurate, imply a portion of the election outcome. 
The paper also presents an interesting example showing that the rules of STV tabulation used for the Australian Senate can allow bizarre behaviour, such as votes increasing in value over time.

\end{abstract}

\section{Introduction}

Single Transferable Vote (STV) is a multi-winner preferential election scheme used widely in Australia to elect candidates to the upper houses of Parliament (such as the Australian Senate). It is also used in Australia for local council elections, in Scotland for local council elections, in the United States, and in Ireland for presidential elections. Recent work has examined the potential of post-election \textit{risk-limiting audits} (RLAs) for STV contests \cite{blomSTV22,blom2024rlas}. These post-election audits aim to provide a certain degree of confidence in the correctness of a reported outcome, where that outcome has been established by machines and computers. For Australian Senate elections, paper ballots cast by votes are scanned to produce digital representations that feed into tabulation software. 

RLAs for STV are hard. While some work has considered 2-seat STV contests \cite{blomSTV22,blom2024rlas}, and has shown that these contests can be audited via an RLA in certain circumstances, no methods are currently available to audit the large STV contests that take place in Australian states. This technical report considers a different question -- whether there are properties of the ballots cast in such STV contests that (i) could conceivably be verified by hand and (ii) if verified, establish the correctness of a portion of the reported election outcome. The properties  considered are the first preference tallies of each candidate--the number of ballots that rank the candidate first--and the number of ballots in each candidates' first preference tally that fall into the categories of \textit{above-the-line} and \textit{below-the-line} votes. In Australian Senate Elections, voters can provide a ranking over available candidates by either ranking parties (or groups) of candidates (above-the-line) or by directly ranking individual candidates (below-the-line). These two types of votes are described in Section \ref{sec:Background}.

Tabulation of STV elections proceeds in rounds of candidate election and elimination. A candidate is elected to a seat when their tally reaches or exceeds a certain threshold of the total valid vote. This threshold is called the election's \textit{quota}. When no candidate has a quota, the candidate with the smallest tally is eliminated. The ballots in their tally pile are distributed to later preferences on the ballot. When a candidate has a quota, they are elected. The ballots in their tally pile are then re-weighted\footnote{Each ballot starts with a value of 1 vote.} before being distributed to later preferences on the ballot. When we have filled all available seats, or the number of candidates left standing equals the number of seats that remain to be filled, we stop. In the latter case, these candidates are elected to a seat. We describe this tabulation process in greater detail in Section \ref{sec:Background}.

This technical report shows that if we treat as accurate the first preference tallies, and above-the-line to below-the-line split of ballots in these tallies, we can generally establish a number of \textit{guaranteed seatings} in the first few rounds of tabulation. While Australian Senate ballots are scanned by machines and counted by software, some counts are performed by hand on election night. At each polling location, the first preference tallies for each party or group is counted by hand\footnote{https://www.aec.gov.au/Voting/counting/senate.htm}. A party's first preference tally includes all ballots that rank the party first in an above-the-line fashion, or a candidate in the party first below-the-line. It not unreasonable to assume that a hand count of each individual candidate's first preference tally, and a count of the number of above-the-line and below-the-line ballots in these tallies, would be feasible. 

The remainder of this report is structured as follows.
Section \ref{sec:Background} describes STV tabulation, following the rules of the Australian Senate, with formal definitions provided in Section \ref{sec:Prelims}. Equations for computing bounds on the tallies of candidates in each round of STV tabulation are presented in Section \ref{sec:Bounds}. The outcomes of Australian Senate Elections between 2016 and 2022 are analysed in Section \ref{sec:Analysis}, using these equations to establish a number of guaranteed seatings. Section \ref{sec:Conclusion} concludes. Acknowledgements can be found in Section \ref{sec:Acks}.

\section{Background}\label{sec:Background}

In an Australian Senate Election, each state and territory hold an STV election to elect between two to twelve candidates to the Australian Senate. Territories elect two candidates to the 76 member chamber. Each state, in a typical election cycle, will elect six candidates. In a double dissolution election, the full twelve seats allocated to each state are up for re-election. In Australia, there are two territories and six states. 

\begin{figure}[t]
\includegraphics[width=\textwidth]{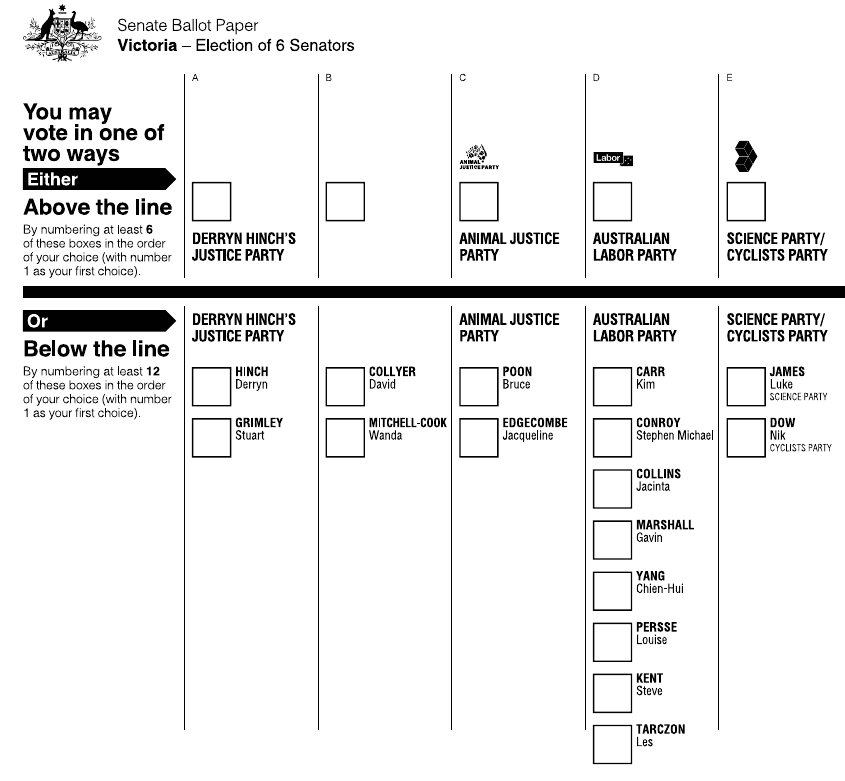}
\caption{Segment of a ballot paper for the 2016 Australian Senate Election for the State of Victoria. The full ballot paper contained 40 groups (columns) and 116 candidates. Source: Wikimedia Commons.}
\label{fig:BallotSample}
\end{figure}

Each ballot cast in an STV election defines a partial or total ranking over the set of available candidates. Figure \ref{fig:BallotSample} shows a segment of a ballot paper for the 2016 Australian Senate Election for the State of Victoria. Each column defines a group of candidates, usually belonging to a specific party. For example, the Victorian ballot paper lists two candidates belonging to the Animal Justice Party (group C). The order of candidates in each of these lists is important. Candidates placed higher in each list will have a greater chance of being elected.  For the Australian Senate, each ballot provides two different ways of ranking candidates -- \textit{above-the-line} or \textit{below-the-line}. 

A voter can provide a preference ranking over the groups (parties) themselves by numbering the boxes that sit \textit{above-the-line}. This ranking will be converted into a ranking over the candidates themselves for the purposes of tabulation. If the voter expresses a first preference for the Animal Justice Party in Figure \ref{fig:BallotSample}, then the two candidates in that group are given the rankings 1 and 2, from the top of the list to the bottom. If the voter then expressed a second preference for the Australian Labor Party, the 8 candidates in that group are given the rankings 3 to 10. Alternatively, the voter can provide a preference ranking over the candidates themselves by numbering the boxes that sit \textit{below-the-line}. 

Tabulation proceeds in rounds of candidate election and elimination. Initially, each candidate is given all ballots on which they are ranked \textit{first}. This represents their \textit{first preference tally}. Each ballot has a value of 1 vote at the start of tabulation. Consider the example STV contest in Table \ref{tab:EGSTV1} involving four candidates ($c_1$ to $c_4$) and two seats. In this election, 55 ballots have been cast. Candidate $c_1$ has a first preference tally of 10 votes, while $c_2$, $c_3$ and $c_4$ have first preference tallies of 15 votes.

The criteria for electing a candidate to a seat is that either they achieve a tally that is equal to or greater than the election's \textit{quota} or they are one of the $R$ candidates left standing with $R$ seats left to fill. The typical formula used to compute an election's quota is provided in Eqn \ref{eqn:Droop} of Definition \ref{def:Election}. In the example of Table \ref{tab:EGSTV1}, the quota is 19 votes. No candidate has a quota's worth of votes on the basis of their first preference tallies. Where no candidate has a quota's worth of votes, the candidate with the smallest tally is \textit{eliminated}. In the example of Table \ref{tab:EGSTV1}, candidate $c_1$ has the smallest tally at 10 votes. The 10 ballots in $c_1$'s tally pile are distributed to the next most preferred \textit{eligible} candidate in their rankings. A candidate is \textit{eligible} to receive votes iff: they are still standing (have not been elected or eliminated); and they do not currently have a quota's worth of votes. An eliminated candidates ballots are distributed to later preferences \textit{at their current value}. The 10 {}[$c_1$, $c_3$] ballots sitting in $c_1$'s tally pile are given to $c_3$ at their current value of 1. Candidate $c_3$ will then have 25 votes in their tally.

When a candidate has at least a quota's worth of votes in their tally at the start of a round, they are elected to a seat. Where multiple candidates have reached a quota as a result of ballot distributions of the prior round, candidates are typically elected in order of their \textit{surplus} (highest to smallest). A candidate's surplus is the difference between their current tally and the quota. Each ballot sitting in the elected candidate's tally pile is re-weighted before it is distributed to the next most preferred eligible candidate. Under the Australian Senate rules, each ballot leaves with the same value--the elected candidate's \textit{transfer value}. This value, defined in Eqn \ref{eqn:tvalue} of Definition \ref{def:TransferValue}, is equal to their surplus divided by the number of ballots in their tally pile. Note that under the Australian Senate rules, a candidate's tally is rounded down to the nearest integer when receiving votes via a surplus transfer. In the example of Table \ref{tab:EGSTV1}, candidate $c_3$ has more than a quota in their tally after receiving ballots from $c_1$. They are elected, and a transfer value of $\tau_2 = (25-19)/25 = 0.24$ computed for round 2. Only the 10 {}[$c_3$, $c_2$, $c_1$] ballots have eligible later preferences, and so these ballots, each with a value of 0.24, are distributed to $c_3$. Candidate $c_3$ now has a tally of 17 votes. 

Continuing the example, there are now two remaining candidates, neither of which has a quota. Candidate $c_4$ has the smallest tally, and is eliminated. The 15 {}[$c_4$, $c_1$, $c_2$] ballots skip $c_1$, who has been eliminated, and are given to $c_2$ at their current value of 1. Candidate $c_2$ is the last remaining candidate, and with one seat left to be filled, is elected. While $c_2$ has amassed a quota at this point, this criteria does not have to be satisfied when we are electing all remaining candidates to fill all remaining seats.

\begin{table}[t]
    \begin{subtable}{.3\columnwidth}
      \centering
        \begin{tabular}{cr}
& \\
& \\
\hline
Ranking & Count \\
\hline
{}[$c_3$, $c_2$, $c_1$] & 10 \\
{}[$c_2$] &  15 \\
{}[$c_4$, $c_1$, $c_2$] & 15 \\
{}[$c_3$, $c_1$] & 5 \\
{}[$c_1$, $c_3$] &  10 \\
\hline
\end{tabular}
        \caption{}
				\label{tab:EGSTV1a}
    \end{subtable}$\,$
    \begin{subtable}{.5\columnwidth}
      \centering
        \begin{tabular}{crrrr}
Seats: 2 & Quota: 19 & & &\\
\hline
Cand. & Round 1 & Round 2 & Round 3 & Round 4\\
\hline
 & $c_1$ eliminated & $c_3$ elected & $c_4$ eliminated & $c_2$ elected\\
& & $\tau_2$ $=$ $0.24$  &  & \\
\hline
{}$c_1$ & 10  & --- & --- & ---\\
{}$c_2$ &  15 & 15 & 17 & 32\\
{}$c_3$ & 15   & 25 & --- & ---\\
{}$c_4$ & 15  & 15 & 15 & ---\\
\hline
\end{tabular}
\caption{}
\label{tab:EGSTV1b}
    \end{subtable} 
    \caption{An STV election profile, stating (a) the number of
ballots cast with each listed ranking over candidates $c_1$ to $c_4$, and (b)
the tallies after each round of counting, election, and elimination. }
\label{tab:EGSTV1}
\end{table}

\section{Preliminaries}\label{sec:Prelims}

\begin{definition}[STV Election] An STV election $\election$ is a tuple $\election = (\cands, \ballots, \quota, \seats)$ where $\cands$ is a set of candidates, $\ballots$ the multiset of ballots cast, $\quota$  the  election quota (the number of votes a candidate must attain to be elected to a seat -- the Droop quota -- Eqn \ref{eqn:Droop}), and $\seats$ the number of seats to be filled.  
\begin{equation}
Q = \Bigl\lfloor \frac{|\ballots|}{N + 1} \Bigr\rfloor + 1
\label{eqn:Droop}
\end{equation} 
\label{def:Election}
\end{definition}

The \textit{value} of ballot $b \in \ballots$ at the start of round $i$ of tabulation is denoted by $v_i(b)$. Recall that each ballot initially has a value of 1, but may be re-weighted during tabulation. 

\begin{definition}[Tally $t_i(c)$] The tally of a candidate $c \in \cands$ in round $i$ is the sum of the values of the ballots in $c$'s tally pile. These are the ballots for which $c$ is ranked first among the set of candidates still standing, $S_i$. Let $\ballots_{i,c}$ denote the subset of ballots sitting in $c$'s tally pile at the start of round $i$.
\begin{equation}
t_i(c) = \lfloor \sum_{b \in \ballots_{i,c}} v_i(b) \rfloor
\end{equation}
Note that under Australian Senate rules, candidate tallies are rounded down to the nearest integer (i.e., when candidates are given ballots via a surplus transfer).
\end{definition}

\begin{definition}[Transfer value $\tau_i$] A transfer value is a re-weighting to apply to ballots being distributed as part of a surplus when a candidate $c$ is elected in round $i$. Let $\ballots_{i,c}$ denote the set of ballots sitting in $c$'s tally pile at the start of round $i$.  The transfer value for round $i$ is computed by dividing $c$'s surplus by the total number of ballots sitting in $c$'s tally pile.
\begin{equation}
\tau_i = \frac{t_i(c) - \quota}{|\ballots_{i,c}|}
\label{eqn:tvalue}
\end{equation} 
\label{def:TransferValue}
\end{definition}

Note that the above definition of a transfer value is specific to the variant of STV used for the Australian Senate. 

\section{Establishing Tally Bounds and Guaranteed Seatings}\label{sec:Bounds}

In a typical Australian Senate election, for a given state or territory,  the outcome of the first few rounds of counting can often be inferred based solely on the first preference tallies of candidates and the proportion of these tallies made up of above-the-line (ATL) and below-the-line (BTL) ballots. We do not need to know what later preferences exist on an ATL ballot to know how the ballot will move between candidates in early rounds of counting. As described in Section \ref{sec:Prelims}, a ballot that ranks a given party or group of candidates first will sit in the first preference pile of the first listed  candidate of that party or group. If that candidate is elected or eliminated, the ballot will move to the next available candidate in that party or group. Once the ballot has reached the last member of the party or group, we would need to know which party was ranked second to determine its next destination. We cannot make similar inferences for BTL ballots. Once the first ranked candidate is elected or eliminated, we need to know the later rankings to determine its next destination. 

The majority of ballots cast by voters in Australian Senate, and other Australian upper house elections, are ATL. The outcome of many of these elections also follows a common pattern. In this pattern, a number of candidates are awarded a seat in the first few rounds of counting, followed by a substantial number of eliminated candidates before the final few seats are awarded.  The analysis in this paper is ultimately more conservative than what is done in practice. A commentator would look at the number of quota's worth of votes a party has received in a given state or territory, by summing the first preference tallies of their candidates, and use that alongside assumptions around preference flows to determine how many seats the party will likely get. Interestingly, one may think that when a party receives at least $n$ quotas worth of votes, they  would win $\lfloor n \rfloor$ seats. See Example \ref{eg:NQuotasNotNSeats} for a scenario where this does not hold. In fact, it is even possible to derive an example where a party that receives $n$ quotas worth of votes does not even receive $n-1$ seats!

\begin{example}
Consider an STV contest with 5 seats and 6 candidates, $a_1 \ldots a_4$ from party $A$, $b$ from party $B$, and $c$ from party $C$\footnote{Thanks to Andrew Conway for providing this example.}. A total of 410 ATL ballots are cast with a first preference for party $A$. A total of 101 BTL ballots are cast with the ranking $[b, a_3, c]$, and 87 ballots with a first preference for $c$. The quota for this contest is 100 votes. Candidates $a_1,$ $b,$ $a_2,$ $a_3$, and $c$ are elected, but not $a_4$ even though party $A$ received over 4 quotas worth of first preference votes. 

Initially, $a_1$, $b$ and $c$ have 410, 101, and 87 first preference votes, respectively, while all other candidates have a first preference tally of 0. Candidates $a_1$ and $b$ have a quota and are elected to a seat, in that order. When $a_1$ is elected, their transfer value is $(410 - 100)/410 = 0.756$. Each ballot leaving their tally will have a value of 0.756. Candidate $a_2$ gets 410 ballots each valued at 0.756, adding 310 votes to their tally. While this gives $a_2$ a quota, and the highest tally, $b$ is the next to be seated as they earned their seat in an earlier round. Candidate $b$'s transfer value is $(101-100)/101 = 0.0099$. All of $b$'s ballots move to $a_3$, with a combined value of 1. Note that while $a_3$ now has a tally of 1, 101 ballots are sitting in their pile. Candidate $a_2$ is now elected, with a transfer value of $(310-100)/410 = 0.512$. The 410 ballots sitting in $a_2$'s pile move to $a_3$ with a combined value of 210. Candidate $a_3$ now has 511 ballots in their tally pile with a combined value of 211 votes. Candidate $a_3$ has a quota and is elected to a seat. Their transfer value is computed as $(211-100)/511 = 0.217$. 

This is now where things get interesting, and rather strange. Every ballot leaving $a_3$ is now \textit{re-weighted} to the transfer value of 0.217. Even those that arrived in $a_3$'s tally with a vale of 0.0099 each! A total of 410 ballots (total value of 89 votes) go to $a_4$. A total of 101 ballots (total value of 22 votes) go to $c$. This gives $c$ a quota at 108 votes, and the final seat.
 
This example, and other interesting edge cases, can be found at:
\begin{center}
    \url{https://github.com/AndrewConway/ConcreteSTV/tree/main/examples}
\end{center}

\begin{table}
\begin{subtable}{.27\columnwidth}
\centering
 \begin{tabular}{cr}
\hline
Ranking & Count \\
\hline
{}[$a_1$, $a_2$, $a_3$, $a_4$] & 410 \\
{}[$b$, $a_3$, $c$] &  101 \\
{}[$c$] & 87 \\
\hline
\end{tabular}
\caption{}
\label{tab:EGSTV2a}
\end{subtable}
\begin{subtable}{0.69\columnwidth}
\begin{tabular}{r|rrrrr}
\multicolumn{6}{l}{Seats: 5 $\quad$ Quota: 100 }\\
\hline
Cand. & Round 1 & Round 2 & Round 3 & Round 4 & Round 5 \\
\hline
 & $a_1$ elected & $b$ elected & $a_2$ elected & $a_3$ elected & $c$ elected \\
& $\tau_1 = 0.756$ & $\tau_2 = 0.0099$  & $\tau_3 = 0.512$ & $\tau_4 = 0.217$ & \\
$a_1$ & {\color{RoyalPurple} 410} & -- & -- & --  & -- \\
$a_2$ & 0   & {\color{RoyalPurple} 310}& {\color{RoyalPurple} 310}& --  & -- \\
$a_3$ & 0   & 0  & 1  & {\color{RoyalPurple}211} & -- \\
$a_4$ & 0   & 0  & 0  & 0   & 89  \\
$b$   & {\color{RoyalPurple} 101} & {\color{RoyalPurple}101}& -- & --  & --  \\ 
$c$   & 87  & 87 & 87 & 87  & {\color{RoyalPurple}108} \\ 
\end{tabular}
\caption{}
\label{tab:EGSTV2b}
\end{subtable}
\caption{\label{tab:FourQuotasDoesntMeanFourSeats_AEC2013} An STV contest with 6 candidates and 5 seats. Elected: $a_1$,
$b$, $a_2$, $a_3$, $c$. Elected candidates are colored
purple once they have a quota.}
\end{table}

\label{eg:NQuotasNotNSeats}
\end{example}

If we accept the first preference tallies of each candidate, and the number of ATL and BTL ballots in each candidates' first preference tally, as accurate, we can establish a number of \textit{guaranteed seatings} in the first few rounds of counting. 
To establish these guaranteed seatings, we can maintain a running lower and upper bound on: the total \textit{value} of ATL and BTL ballots in a candidates tally; and the total \textit{number} of ATL and BTL ballots in a candidates tally. This paper defines a number of equations for computing these lower and upper bounds  in each round of tabulation of an STV contest. These equations make use of the fact that we know how a ballot with an above-the-line ranking will move between the candidates in the first ranked group. While commentators will use predictions about the later preferences on ballots, our analysis does not make any assumptions on where a ballot sitting in the first preference tally of a candidate $c$ will move to when $c$ is either elected or eliminated, \textit{unless} the ballot contains an above-the-line vote and $c$ is not the last candidate in their group.\\

\noindent The following additional notation is used:  

\begin{itemize}
\setlength\itemsep{1em}
\item $\underline{t}^{ATL}_{i,c}$ and $\underline{t}^{BTL}_{i,c}$ denote \textit{lower bounds} on the total value of ATL and BTL ballots, respectively, in candidate $c$'s tally at the start of round $i$. 

\item $\overline{t}^{ATL}_{i,c}$ and $\overline{t}^{BTL}_{i,c}$ denote \textit{upper bounds} on the total value of ATL and BTL ballots, respectively, in candidate $c$'s tally at the start of round $i$.

\item  $\underline{p}^{ATL}_{i,c}$, $\underline{p}^{BTL}_{i,c}$, 
$\overline{p}^{ATL}_{i,c}$ and $\overline{p}^{BTL}_{i,c}$ denote lower and upper bounds on the number of ATL and BTL ballots in a candidate $c$'s tally at the start of round $i$. Note that these bounds relate to pieces of paper, and not vote tallies.

\item $\underline{t}_{i,c}$ and $\overline{t}_{i,c}$ denote lower and upper bounds on the \textit{total tally} of candidate $c$ at the start of round $i$.

\item $\underline{p}_{i,c}$ and $\overline{p}_{i,c}$ denote lower and upper bounds on the \textit{total number of ballots} (pieces of paper) in the tally of candidate $c$ at the start of round $i$.

\item $\underline{s}_{i}$ and $\overline{s}_i$ denote lower and upper bounds on the surplus of the candidate elected in round $i$.

\item $\underline{\tau_i}$ and $\overline{\tau_i}$ denote lower and upper bounds on the transfer value assigned to ballots leaving the elected candidate's tally in round $i$.
\end{itemize}

In the first round of counting, $i = 1$, the lower and upper bounds on tallies (both ATL, BTL, and total) are equal. This is a result of assuming that we have accurate first preference tallies, and an accurate count of how many ballots in each candidate's tally pile contain ATL and BTL votes.\\

\noindent For all candidates $c \in \cands$:

\begin{eqnarray}
\underline{t}^{ATL}_{i=1,c} & \equiv \overline{t}^{ATL}_{i=1,c}  \\
\underline{t}^{BTL}_{i=1,c} & \equiv \overline{t}^{BTL}_{i=1,c} \\ 
\underline{t}_{i=1,c} & \equiv \overline{t}_{i=1,c} \\
\underline{p}^{ATL}_{i=1,c} & \equiv \overline{p}^{ATL}_{i=1,c} \\
\underline{p}^{BTL}_{i=1,c} & \equiv \overline{p}^{BTL}_{i=1,c} \\ 
\underline{p}_{i=1,c} & \equiv \overline{p}_{i=1,c} 
\end{eqnarray}

 The following equations compute upper and lower bounds on candidate tallies as candidates are elected and eliminated. The resulting bounds are not informative after you have reached the first elimination--they become too loose. In cases where several candidates are elected in the first few rounds, however, the bounds tell us that some of those elections are guaranteed if we assume the first preference tallies and the ATL to BTL split in those tallies are correct. Sections \ref{sec:FG_Elections} and \ref{sec:FG_Eliminations} outline how these bounds are updated after each election and elimination of a candidate.  Note that any bound on a candidates' tally, or the number of papers in their tally, cannot exceed the total number of ballots cast in the election. Bounds on ATL tallies, and the number of ATL ballots in a candidates' tally, cannot exceed the total number of ATL ballots cast in the election. 

\subsection{Elections}\label{sec:FG_Elections}
When a candidate $e$ is elected in round $i$, the upper and lower tally bounds for remaining candidates are updated as follows. First, a minimum and maximum transfer value on the basis of the elected candidate's minimum and maximum tally is computed. The minimum and maximum tally for $e$ in round $i$, $\underline{t}_{i,e}$ and $\overline{t}_{i,e}$, is computed as per Eqns \ref{eqn:minvotes} and \ref{eqn:maxvotes}. The minimum and maximum number of ballots in a candidate $e$'s tally in round $i$, $\underline{p}_{i,e}$ and $\overline{p}_{i,e}$, is computed as per Eqns \ref{eqn:minpapers} and \ref{eqn:maxpapers}. 

\begin{eqnarray}
    \underline{t}_{i,e} & = & \underline{t}^{ATL}_{i,e} + \underline{t}^{BTL}_{i,e} \label{eqn:minvotes}\\
    \underline{p}_{i,e} & = & \underline{p}^{ATL}_{i,e} + \underline{p}^{BTL}_{i,e} \label{eqn:minpapers}\\
    \overline{t}_{i,e} & = & \overline{t}^{ATL}_{i,e} + \overline{t}^{BTL}_{i,e} \label{eqn:maxvotes}\\
    \overline{p}_{i,e} & = & \overline{p}^{ATL}_{i,e} + \overline{p}^{BTL}_{i,e} \label{eqn:maxpapers}
\end{eqnarray}
To compute minimum and maximum transfer values for the elected candidate $e$ in round $i$, we first compute minimum and maximum surpluses.
\begin{eqnarray}
    \underline{s}_{i} &=& \max(0, \underline{t}_{i,e} - \mathcal{Q}) \\ 
    \overline{s}_{i} &=& \max(0, \overline{t}_{i,e} - \mathcal{Q}) 
\end{eqnarray}
The minimum and maximum transfer value  for the elected candidate $e$ is then:
\begin{eqnarray}    
\underline{\tau_i} &=& \min(1,  \frac{\underline{s}_{i}}{\overline{p}_{i,e}})\\
\overline{\tau_i} & = & \min(1, \frac{\overline{s}_{i}}{\underline{p}_{i,e}}) 
\end{eqnarray}
First, we assume that candidate $e$'s BTL ballots could move to any remaining candidate $c$ whose minimum tally in round $i$ is less than a quota.
\begin{eqnarray}
\overline{t}^{BTL}_{i+1,c} &= & \overline{t}^{BTL}_{i,c} + \overline{\tau_i} \, \overline{p}^{BTL}_{i,e} \\
\overline{p}^{BTL}_{i+1,c} & = &  \overline{p}^{BTL}_{i,c} + \overline{p}^{BTL}_{i,e}
\end{eqnarray}
If candidate $e$ is part of a group of candidates, and is not the \textit{last} listed candidate, all of their ATL ballots will move to the next listed candidate in their group. Let's call this candidate $n$. Note that as candidate tallies are rounded down to the nearest integer upon the transfer of surpluses, we round down the contribution of the surplus transfer in Eqn \ref{eqn:LBTallySurplus} when computing the lower bound on candidate $n$'s ATL tally.   
\begin{eqnarray}
    \overline{t}^{ATL}_{i+1,n} & = \overline{t}^{ATL}_{i,n} + \overline{\tau_i} \, \overline{p}^{ATL}_{i,e} \\
    \overline{p}^{ATL}_{i+1,n} & = \overline{p}^{ATL}_{i,n} + \overline{p}^{ATL}_{i,e} \\
    \underline{t}^{ATL}_{i+1,n} & = \underline{t}^{ATL}_{i,n} + \lfloor\underline{\tau_i} \, \underline{p}^{ATL}_{i,e}\rfloor \label{eqn:LBTallySurplus}\\
    \underline{p}^{ATL}_{i+1,n} & = \underline{p}^{ATL}_{i,n} + \underline{p}^{ATL}_{i,e} 
\end{eqnarray}
If candidate $e$ is the last listed candidate in their group, we do not know where their ATL ballots will move to next without knowledge of the actual rankings on those ballots. Consequently, those ballots could move to any remaining candidate whose minimum tally is less than a quota in round $i$. These ballots are treated as if they were BTL ballots moving forward, further incrementing the maximum BTL tally and paper totals for remaining candidates.   
\begin{eqnarray}
\overline{t}^{BTL}_{i+1,c} &\pluseq & \overline{\tau_i} \, \overline{p}^{ATL}_{i,e} \\
\overline{p}^{BTL}_{i+1,c} & \pluseq &  \overline{p}^{ATL}_{i,e}
\end{eqnarray}
Note that the lower bounds on the BTL vote and ballot paper totals are never incremented. Without knowing the actual ranking on these ballots, we must assume that the ballots may exhaust without being transferred to another candidate.

\subsection{Eliminations}\label{sec:FG_Eliminations}

When a candidate $e$ is eliminated in round $i$, we assume that any remaining candidate $c$ could receive their BTL ballots. 
\begin{eqnarray}
\overline{t}^{BTL}_{i+1,c} &= & \overline{t}^{BTL}_{i,c} + \overline{t}^{BTL}_{i,e} \\
\overline{p}^{BTL}_{i+1,c} & = &  \overline{p}^{BTL}_{i,c} + \overline{p}^{ATL}_{i,e}\\
\end{eqnarray}
If the candidate $e$ is not the last candidate in their group, the ATL ballots in their tally pile will move to the next eligible candidate in the group, $n$.
\begin{eqnarray}
    \overline{t}^{ATL}_{i+1,n} = \overline{t}^{ATL}_{i,n} +   \overline{t}^{ATL}_{i,e} \\
    \overline{p}^{ATL}_{i+1,n} = \overline{p}^{ATL}_{i,n} + \overline{p}^{ATL}_{i,e} \\
    \underline{t}^{ATL}_{i+1,n} = \underline{t}^{ATL}_{i,n} +  \underline{t}^{ATL}_{i,e} \\
    \underline{p}^{ATL}_{i+1,n} = \underline{p}^{ATL}_{i,n} + \underline{p}^{ATL}_{i,e} \\
\end{eqnarray}
Otherwise, any ATL ballots in their tally pile are treated like BTL ones -- we assume they could move to any remaining candidate $c$.
\begin{eqnarray}
\overline{t}^{BTL}_{i+1,c} &\pluseq &  \overline{t}^{ATL}_{i,e} \\
\overline{p}^{BTL}_{i+1,c} & \pluseq &  \overline{p}^{ATL}_{i,e}
\end{eqnarray}

\subsection{Guarantees}\label{sec:Guarantees}

If a candidate $c$ has a lower bound on their tally that is equal to or greater than a quota in any round, then they are \textit{guaranteed} to be elected.

\section{Analysis}\label{sec:Analysis}

\begin{table}[t!]
    \centering
    \begin{tabular}{llll}
    \toprule
        State/Territory &  $|\mathcal{C}|$ & Seats & Pattern \\
             \hline
     \multicolumn{4}{l}{2022} \\
     \midrule
     VIC & 79 & 6 & \underline{{\bf q q q q}} e $\ldots$ q e $\ldots$ s  \\
     WA & 58 & 6  & \underline{{\bf q q q}} q e $\ldots$ q e $\ldots$ q\\
     TAS & 39 & 6 & \underline{{\bf q q q}} e $\ldots$ q e e q e e q e \\
     SA & 51 & 6 & \underline{{\bf q q q q}} e $\ldots$ q e e e e q \\
     QLD & 79  & 6 & \underline{{\bf q q q}} e $\ldots$ q e e q e q \\
     NSW & 75 & 6 & \underline{{\bf q q q q}} e $\ldots$ q e e e q  \\
     NT & 17 & 2 & e $\ldots$ q e q $\ldots$  \\
     ACT & 23 & 2 & e $\ldots$ q e $\ldots$ q $\ldots$  \\
     \midrule
     \multicolumn{4}{l}{2019} \\
     \midrule
      VIC  &  82  &  6  &  \underline{{\bf q q q q}} e $\ldots$ q e e q  \\  
      WA   &  67  &  6  & \underline{{\bf q q q}} e $\ldots$ q e q e e q $\ldots$ \\
      TAS &  44   &  6  & \underline{{\bf q q}} q e $\ldots$ q e e e q q $\ldots$\\
      SA &   42   &  6  & \underline{{\bf q q q q}} e $\ldots$ q e e q $\ldots$  \\
      QLD &   83  &  6  & \underline{{\bf q q q}} e $\ldots$ q q e q \\
      NSW &  105   &  6  & \underline{{\bf q q q q}} e $\ldots$ q q \\
      NT  &   18  &   2 & \underline{{\bf q q}} $\ldots$  \\
      ACT &   17  &   2 & \underline{{\bf q}} e $\ldots$ q $\ldots$ \\
    \midrule
     \multicolumn{4}{l}{2016} \\
     \midrule
           VIC  &   116 &  12  & \underline{{\bf q q q q q q q}} q e $\ldots$ q e $\ldots$ q e e e q s  \\  
      WA   &  79  &  12  & \underline{{\bf q q q q q q q q}} e $\ldots$ q e $\ldots$  q e q q \\
      TAS &   58  &  12 &  \underline{{\bf q q q q q}}  q q q e $\ldots$ q q q e q \\
      SA &   64   &  12  &  \underline{{\bf q q q q q}} q q q q e $\ldots$ q e q e q \\
      QLD &  122   &  12  & \underline{{\bf q q q q q}} q q q e $\ldots$ q e $\ldots$ q e q s \\
      NSW &  151   &  12  & \underline{{\bf q q q}} q q q q q e $\ldots$ q e $\ldots$ q q s \\
      NT  &  19   &   2 & \underline{{\bf q q}} $\ldots$   \\
      ACT &  22   &   2 & \underline{{\bf q}} e $\ldots$ q $\ldots$\\
      \bottomrule
    \end{tabular}
    \caption{Seating and elimination patterns arising across the Australian Senate elections between 2016 and 2022, across states and territories. A `q' indicates that a candidate has been seated with a quota, a `s' indicates that a candidate has been seated without a quota, and an `e' that one has been eliminated. A `$\ldots$' either represents a string of eliminations, or when present at the tail of a pattern, that there are one or more candidates that remain standing when the last seat has been awarded. A prefix of seatings that are guaranteed if we accept the first preference tallies for each candidate, and the proportion of those tallies that are made up of ATL and BTL ballots, are both in bold and underlined.}
    \label{tab:Guarantees}
\end{table}

Table \ref{tab:Guarantees} records the pattern of seatings and eliminations that occurred in the reported outcome of each Australian Senate election between 2016 and 2022. A `q' indicates that a candidate has been seated with a quota, a `s' indicates that a candidate has been seated without a quota, and an `e' that one has been eliminated. A `$\ldots$' either represents a string of eliminations, or when present at the tail of a pattern, indicates that there are one or more candidates that remain standing when the last seat has been awarded. A prefix of seatings that is guaranteed if we accept the first preference tallies for each candidate, and the proportion of those tallies that are made up of ATL and BTL ballots, is both in bold and underlined.

The sequence  q q q q e $\ldots$ q e e q for Victoria in 2019 indicates that: four candidates were awarded a seat in the first four rounds; a series of candidates were then eliminated; a candidate was seated when there were four candidates left standing; two candidates were then eliminated, leaving one candidate standing to take the last seat (who happened to also have achieved a quota with the last elimination). The sequence q e $\ldots$ q $\ldots$ for ACT in 2016 indicates that the first of two seats was awarded based on first preferences, followed by a string of eliminations, and a seating. The trailing $\ldots$ indicates that at the point of the last seating, one or more candidates remained standing, neither seated or eliminated.  

Consider SA 2016. We can say that the first five candidates elected \textit{are guaranteed} to be the first five candidates elected to a seat. After this point, it is possible that a candidate not elected in the initial tranche of seatings, according to the reported results, could conceivably be elected next under some realisation of  ballot preferences (that has the same assumed first preference tallies and the same assumed proportion of ATL and BTL ballots in those tallies).

\section{Conclusion}\label{sec:Conclusion}

This paper presents a series of equations for computing lower and upper bounds on candidate tallies across rounds of tabulation in an Australian Senate STV contest. These equations assume the correctness of several properties: the first preference tally counts for each candidate; and the number of ATL to BTL ballots in each candidates' first preference tally pile. As first preference tallies are counted by hand on election night in such contests, we could expect these properties to be verified manually. Using these bound calculations, we can show that in many Australian Senate contests the first handful of seatings are `guaranteed' under the assumption that these properties are accurate. 

This knowledge could be used when auditing large STV contests like those held for the Australian Senate. The audit may want to assume the correctness of these initial seatings, and concentrate on verifying later ones. We know from prior work on RLAs for STV \cite{blomSTV22,blom2024rlas} that assumptions on some of the candidates that win, and when they are seated, are useful when verifying later winners.

\section{Acknowledgements}\label{sec:Acks}

I would like to acknowledge the helpful comments of Vanessa Teague, Peter Stuckey, Damjan Vukcevic, Alexander Ek, and Andrew Conway. A special thanks is due to Andrew Conway for developing the contest in Example \ref{eg:NQuotasNotNSeats} showing how ballots in an Australian Senate contest can increase in value over time. 

\bibliographystyle{splncs04}
\bibliography{bib}
\end{document}